\documentclass[twocolumn,prd,superscriptaddress,preprintnumbers,amsmath,amssymb,nofootinbib]{revtex4}
\newcommand{\bg}{\bar{g}}
\usepackage{amsmath}
\usepackage{amssymb}
\usepackage{slashed}
\usepackage{graphicx}
\usepackage{graphics}
\usepackage{epsfig}
\usepackage{color}
\usepackage{hyperref}
\newcommand{\Eqref}[1]{eq.~\eqref{#1}}

\begin{document}
\title{Faddeev-Popov ghosts in quantum gravity beyond perturbation theory}

\author{Astrid Eichhorn}
\affiliation{\mbox{\it Perimeter Institute for Theoretical Physics, 31 Caroline Street N, Waterloo, N2L 2Y5, Ontario, Canada}
\mbox{\it E-mail: {aeichhorn@perimeterinstitute.ca}}}

\begin{abstract} 
We study the Faddeev-Popov ghost sector of asymptotically safe quantum gravity, which becomes non-perturbative in the ultraviolet. We point out that nonzero matter-ghost couplings and higher-order ghost self-interactions exist at a non-Gau\ss{}ian fixed point for the gravitational couplings, i.e., in the ultraviolet. Thus the ghost sector in this non-perturbative ultraviolet completion does not keep the structure of a simple Faddeev-Popov determinant.
We discuss implications of the new ghost couplings for the Renormalization Group flow in gravity, the form of the ultraviolet completion, and the relevant couplings, i.e., free parameters, of the theory.

\end{abstract}

\maketitle

\section{Introduction}
Asymptotically  safe quantum gravity has been studied in various approximations of the effective action, for reviews see \cite{Nagy:2012ef,Reuter:2012xf,Reuter:2012id,Percacci:2011fr,Litim:2011cp,Litim:2008tt,Percacci:2007sz,Niedermaier:2006ns,Niedermaier:2006wt}: Starting from an Einstein-Hilbert term \cite{Reuter:1996cp}, see also \cite{Litim:2003vp,Fischer:2006fz}, curvature squared \cite{Lauscher:2002sq,Codello:2006in} and higher-order scalar curvature truncations \cite{Codello:2008vh,Benedetti:2012dx,Dietz:2012ic} and more complicated tensor structures \cite{Benedetti:2009rx} have been studied. A setting with Lorentzian signature for the quantum fluctuations has been investigated \cite{Manrique:2011jc,Rechenberger:2012dt}, and other choices of fundamental variables have been explored \cite{Daum:2010qt,Harst:2012ni, Dona:2012am}. A connection to the semiclassical regime in the infrared (IR) has been established \cite{Reuter:2004nx}, with indications for a possible IR fixed point \cite{Donkin:2012ud,Nagy:2012rn,Christiansen:2012rx}.
The study of the Faddeev-Popov ghost sector has been initiated in \cite{Eichhorn:2009ah}, see also \cite{Eichhorn:2011gc}, and continued in \cite{Eichhorn:2010tb,Groh:2010ta}, and the bimetric structure arising from the gauge-fixing term has also been studied \cite{Manrique:2009uh,Manrique:2010mq,Manrique:2010am}. So far, all the studies apart from \cite{Eichhorn:2009ah} assume a simple structure of the Faddeev-Popov ghost sector:
The usual gauge-fixing procedure in gauge-theories such as quantum gravity in the path-integral framework employs the Faddeev-Popov trick, which results in the Faddeev-Popov (FP) determinant in the generating functional. Using Grassmann-valued fields, this determinant can be exponentiated, yielding a local action with dynamical ghost fields. Standard choices of gauge fixing, such as the harmonic gauge, yield a ghost action which is quadratic in the ghosts.
Beyond the perturbative regime, this structure changes: Metric fluctuations induce further terms beyond a simple FP ghost sector. Here, we will focus on the existence of ghost-matter interactions as well as higher-order ghost self-interactions. These are usually not present in gauge theories in the ultraviolet (UV), since they do not arise from the perturbative FP trick. In the case of asymptotic safety, where the theory becomes non-perturbative in the UV, such terms are generated by metric fluctuations and will be non-zero at the UV interacting fixed point.
This implies the following structure of the ghost sector in the UV: Due to the existence of higher-order ghost operators, it is not possible to straightforwardly reverse the FP trick. In the case of an asymptotically safe gauge theory, the ghost sector seems to be part of the very definition of the microscopic action.
The existence of these couplings also raises the question how possible relevant couplings in the ghost sector should be understood, and whether the status of the Gribov problem differs fundamentally between asymptotically free and asymptotically safe gauge theories. 

We will show that ghost-antighost-2-scalar interactions and fourth-order ghost terms are generated by the Renormalization Group (RG) flow, as soon as a kinetic term for the scalar matter and the standard Faddeev-Popov ghost term are present. 
In fact these are only the first terms in what is to be expected an infinite number of new terms with nonzero couplings at the fixed point.

\begin{figure}[!here]
\includegraphics[width=0.4\linewidth]{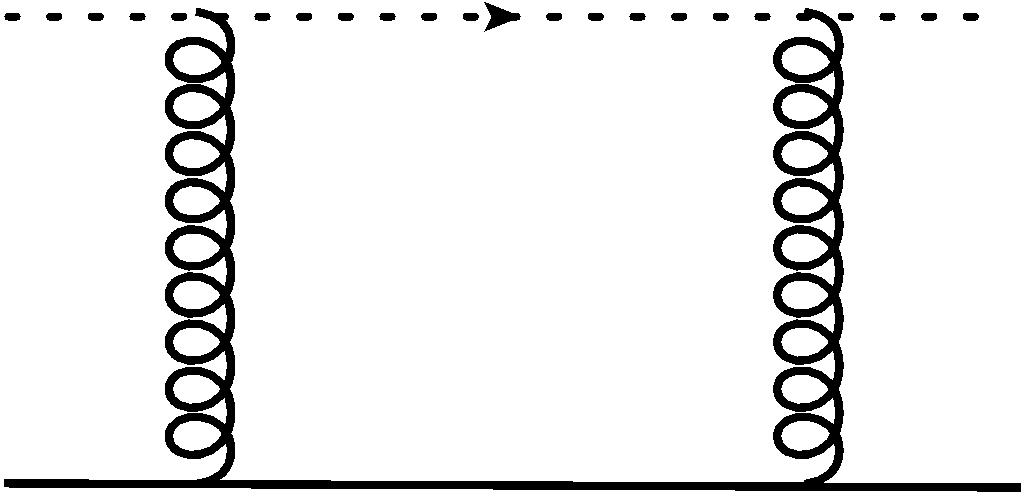} \quad \includegraphics[width=0.2\linewidth]{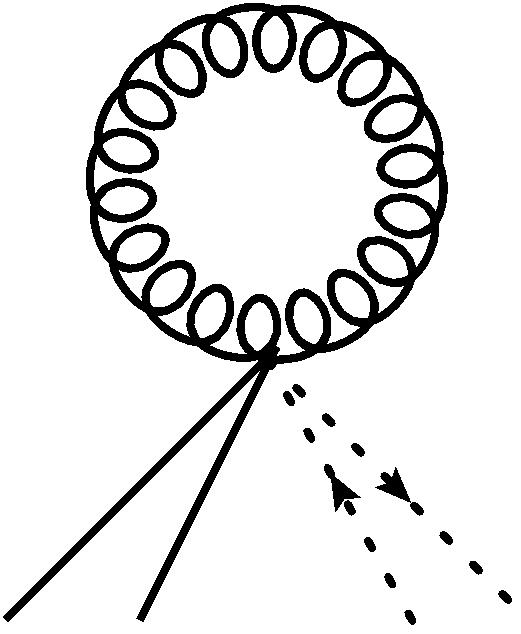}
\caption{\label{ex_diags} These diagrams contribute to the $\beta$ function for a ghost-matter coupling. Dotted lines denote ghosts, spiralling lines metric propagators and full lines denote scalars.}
\end{figure}

In order to show that these new couplings are nonzero, it suffices to evaluate a subset of all terms in their $\beta$ functions.
In general, these $\beta$ functions contain the following types of terms, see fig.~\ref{ex_diags}:

\begin{itemize}
\item Terms which generate these interactions even if they are set to zero at some scale. These contributions are $\sim Z_{i}^n G_N^2$, since they are generated from the kinetic terms only. Herein $Z_i$ denotes the wave function renormalization of the matter and ghost field, respectively, and $n$ is the number of vertices in the respective diagrams. $G_N$ denotes the Newton coupling. For instance, the diagram to the left in fig.~\ref{ex_diags} yields a contribution $\sim G_N^2$, since each metric propagator comes with a factor of $G_N$. The vertices in the diagram arise from the kinetic terms, and are therefore $\sim Z_i$.
\item Further terms are proportional to (powers of) the coupling itself. As an example, the diagram to the right in fig.~\ref{ex_diags} is proportional to the matter-ghost coupling itself, as the vertex is proportional to it.
\end{itemize}

The first type of contribution implies that these couplings cannot approach a Gau\ss{}ian fixed point, i.e., their fixed-point values are necessarily nonzero, as soon as $G_N$ is nonzero. Accordingly these terms induce a shift in the $\beta$ function, such that the Gau\ss{}ian fixed point becomes shifted to an interacting one. The second contribution can induce further non-Gau\ss{}ian fixed points at larger values of the coupling, and also enters the critical exponent at the shifted Gau\ss{}ian fixed point. Here, we will assume that at this fixed point, metric fluctuations yield the dominant contribution to the $\beta$ function. Thus contributions that are proportional to the coupling itself will be subleading, and we will neglect them.
Although our calculation is an approximation to the full $\beta$ functions within our truncation, it suffices to show that the couplings under investigation cannot have a vanishing fixed-point value. This calculation is therefore sufficient to show that the structure of the Faddeev-Popov ghost sector of the microscopic fixed-point action differs crucially from the perturbative regime.

\section{Calculation of ghost-matter couplings and ghost self-interactions}

We will employ a non-perturbative formulation of the functional Renormalization 
Group (FRG), for reviews see \cite{Berges:2000ew,Polonyi:2001se,Pawlowski:2005xe,Gies:2006wv,Rosten:2010vm}.
%The Wetterich equation \cite{Wetterich:1993yh} allows to evaluate $\beta$ functions even in the non-perturbative regime.
 We employ a momentum scale $k$  and an IR mass-like regulator function $R_k(p)$, which suppresses IR modes (with $p^2 <k^2$) in the generating functional. The scale-dependent effective action $\Gamma_k$ then contains the effect of quantum fluctuations above the scale $k$ only, and gives the standard effective action $\Gamma$ for $k \rightarrow 0$. Its scale-dependence is given by the following functional differential equation, the Wetterich equation \cite{Wetterich:1993yh}:
\begin{equation}
\partial_t \Gamma_k= \frac{1}{2} {\rm STr} \left(\Gamma_k^{(2)}+R_k \right)^{-1}\partial_t R_k.
\end{equation}
Herein $\partial_t = k\, \partial_k$, and $\Gamma_k^{(2)}$ is matrix-valued in field space and denotes the second functional derivative of the effective action with respect to the fields. Adding the mass-like regulator and taking the inverse yields the full, momentum- and field-dependent propagator.
The supertrace contains a trace over all indices with a negative sign for Grassmann valued fields. In the case of a continuous momentum variable it implies an integration over the momentum, otherwise the discrete eigenvalues of the full regularized propagator are being summed over with the appropriate degeneracy factors included.
On the technical side, the main advantage of this equation is its one-loop form, since it can be written as the supertrace over the full propagator, with the regulator insertion  $\partial_t R_k$ in the loop. Nevertheless it also yields higher terms in a perturbative expansion, see, e.g., \cite{Reuter:1993kw}, since it depends on the full, field- and momentum-dependent propagator, and not just on the perturbative propagator. Expanding the flow equation in a series of operators compatible with the symmetry then allows to extract the non-perturbative $\beta$ functions of the corresponding couplings.

For reasons of practicality, the full RG flow in the infinite-dimensional theory space cannot be evaluated, even though infinite-dimensional truncations can be studied even in gravity \cite{Benedetti:2012dx,Dietz:2012ic}. Thus theory space is truncated. Here, several ways to proceed are possible: Firstly, one could choose the truncation to be the same on both sides of the flow equation. This amounts to examining the RG flow of a number of couplings, which are driven by quantum fluctuations of precisely the operators corresponding to these couplings. Another possibility is to specify a truncation for the right-hand side of the flow equation, which implies that we fix the spectrum of quantum fluctuations that drive the RG flow. It is then possible to consider a different (in particular larger) set of operators on the left-hand side. In this case we study the RG flow of a number of couplings as driven by a smaller subset. Here, we will focus on this option, since it provides the following interesting information: Specifying a minimal set of couplings that we have identified to be non-vanishing in a certain physical setting, this method allows to check which further couplings will be induced by the minimal set of operators, and whether it is possible to set the couplings in a subspace of theory space to zero consistently. Here, we will show that starting from a minimal ghost sector with a Faddeev-Popov term, further ghost couplings are necessarily generated and cannot be set to zero.

To this end we proceed as follows: Splitting $\Gamma_k^{(2)}+R_k =\mathcal{P}_k+\mathcal{F}_k$, where all scalar-field dependent and ghost-dependent
terms enter the fluctuation matrix $\mathcal{F}_k$, such that $\mathcal{P}_k$ is the propagator which does not depend on the external fields, we may
now expand the right-hand side of the flow equation as
follows:
\begin{eqnarray}
 \partial_t \Gamma_k&=& \frac{1}{2}{\rm STr} \{
 [\Gamma_k^{(2)}+R_k]^{-1}(\partial_t R_k)\}\label{eq:flowexp}\\
&=& \frac{1}{2} {\rm STr}\, \tilde{\partial}_t\ln
\mathcal{P}_k
+\frac{1}{2}\sum_{n=1}^{\infty}\frac{(-1)^{n-1}}{n} {\rm
  STr}\,
\tilde{\partial}_t(\mathcal{P}_k^{-1}\mathcal{F}_k)^n,
\nonumber
\end{eqnarray}
where the derivative $\tilde{\partial}_t$ in the second line by definition
acts only on the $k$ dependence of the regulator, i.e.,  $\tilde{\partial}_t=\int \partial_t R_k\frac{\delta}{\delta R_k}$. Since each
factor of $\mathcal {F}_k$ contains a
coupling to external fields, this expansion simply corresponds to an expansion
in the number of vertices. Thus we can straightforwardly write down the diagrammatic expansion of a $\beta$ function.

In the following we will employ the background field
 formalism \cite{Abbott:1980hw}, where the full metric is
split according to
\begin{equation}
 g_{\mu \nu}= \bar{g}_{\mu \nu}+ h_{\mu \nu}.
\end{equation}
Crucially, this split does not mean that we consider only small fluctuations
around a fixed, e.g., flat background. Within the FRG approach we can access
physics also in the fully non-perturbative regime.  The background-field formalism is used in gravity, because the
background metric allows for a meaningful notion of "high-momentum" and
"low-momentum" modes as implied by the spectrum of the background covariant
Laplacian. Later, we will set the background to be flat for reasons of technical simplicity. Note that the $\beta$ functions are independent of a specific choice of background field configuration -- apart from topological considerations, see, e.g., \cite{Reuter:2008qx,Bonanno:2012dg}-- that allows to uniquely project onto the operators under consideration.

We now perform a York decomposition of the fluctuation field $h_{\mu \nu}$ into a transverse traceless symmetric tensor, a transverse vector, a scalar and the trace, and specialize to Landau deWitt gauge, where only the transverse traceless and the trace mode contribute to the running of ghost self-couplings and ghost-matter couplings.

\subsection{Interactions between ghosts and scalar matter}

We consider the following truncation on the right-hand side of the flow equation and will focus on some of the terms that are induced on its left-hand side.
\begin{equation}
 \Gamma_k = \Gamma_{k \,\rm EH}+ \Gamma_{k\,\rm gf}+ \Gamma_{k\, \rm gh} + \Gamma_{k\, \rm matter},\label{truncation}
\end{equation}
where 
\begin{eqnarray}
\Gamma_{k\,\mathrm{EH}}&=& 2 \bar{\kappa}^2 Z_{\text{N}} (k)\int 
d^4 x \sqrt{g}(-R+ 2 \bar{\lambda}(k))\label{eq:GEH},\\
\Gamma_{k\,\mathrm{gf}}&=& \frac{Z_{\text{N}}(k)}{2\alpha}\int d^4 x
\sqrt{\bar g}\, \bar{g}^{\mu \nu}F_{\mu}[\bar{g}, h]F_{\nu}[\bar{g},h]\label{eq:Ggf},
\end{eqnarray}
with
\begin{equation}
 F_{\mu}[\bar{g}, h]= \sqrt{2} \bar{\kappa} \left(\bar{D}^{\nu}h_{\mu
   \nu}-\frac{1+\rho}{4}\bar{D}_{\mu}h^{\nu}{}_{\nu} \right). 
\end{equation}
Here, $\bar{\kappa}= (32 \pi G_{\text{N}})^{-\frac{1}{2}}$ is related to the
bare Newton coupling $G_{\text{N}}$. The standard Faddeev-Popov ghost term reads
\begin{eqnarray}
 \Gamma_{k\, \rm gh}&=& -\sqrt{2} \int d^4x \sqrt{\bg}\, Z_c(k) \,\bar{c}_{\mu} 
\Bigl(\bar{D}^{\rho}\bar{g}^{\mu \kappa}g_{\kappa \nu}D_{\rho}\nonumber\\
&+& \bar{D}^{\rho}\bar{g}^{\mu \kappa}g_{\rho \nu}D_{\kappa}
- \frac{1}{2}(1+\rho)\bar{D}^{\mu}\bar{g}^{\rho \sigma}g_{\rho \nu}D_{\sigma} \Bigr)c^{\nu},
\end{eqnarray}
with a wave-function renormalization $Z_c(k)$.
We now specialize to the Landau deWitt gauge, where $\rho \rightarrow \alpha$ and $\alpha \rightarrow 0$, which is a fixed point of the RG flow.

We will work on a flat background which  suffices to point out the generation of matter-ghost couplings.

\begin{figure}[!here]
\includegraphics[width=0.2\linewidth]{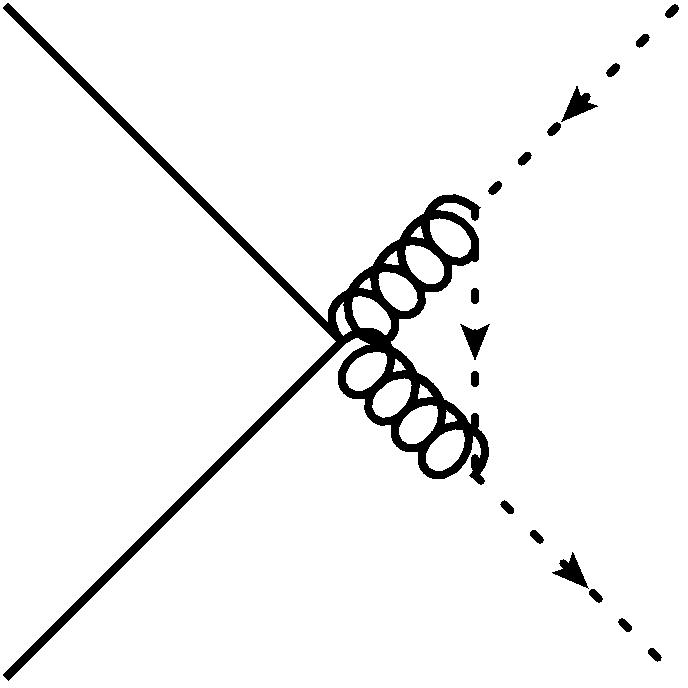}\\
$\phantom{x}$\\
\includegraphics[width=0.4\linewidth]{gh_sc_fourvert1.jpg}\quad 
\includegraphics[width=0.4\linewidth]{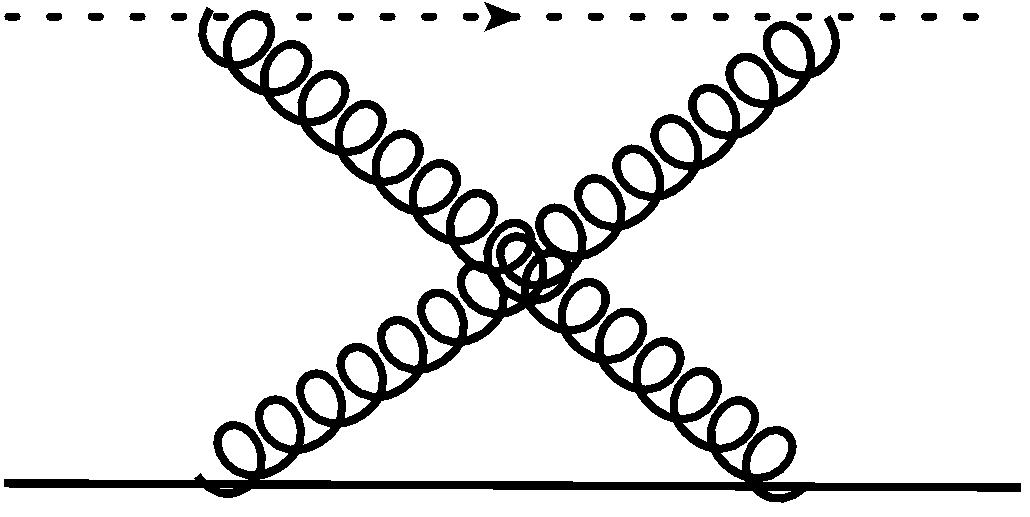}\\
\caption{\label{Gh_sc_diags} These diagrams generate the matter-ghost coupling between $\phi^2$ and a ghost and antighost and thereby remove the Gau\ss{}ian fixed point in the corresponding $\beta$ function. Matter fields are denoted by thick lines, ghosts by dashed and metric fluctuations by spiralling lines. A regulator-insertion exists on each of the internal propagators.}
\end{figure}

We consider minimally coupled scalar matter where
\begin{equation}
\Gamma_{k\, \rm matter}= \frac{Z_{\phi}(k)}{2} \int d^4x \sqrt{g}\, g^{\mu \nu} \partial_{\mu} \phi \,\partial_{\nu} \phi,
\end{equation}
with a wave-function renormalization $Z_{\phi}(k)$.
The flow then generates matter-ghost interactions by the diagrams in fig.~\ref{Gh_sc_diags}.

In order to point out that such interactions are generated, we project the flow onto the following flat-space approximation of the matter-ghost action. The couplings in this induced action are nonzero due to quantum fluctuations, even if set to zero at some scale $k_0$:
\begin{equation}
\Gamma_{k\, \rm ind}=\!\! \int_{p_1,p_2,p_3} \! \! \! \! \! \!\! \! \!  \! \! \! \! V_{\mu \nu}(p_1,p_2,p_3) \bar{c}^{\mu}(p_1) c^{\nu}(p_2) \phi(p_3)\, \phi(p_1-p_2-p_3),
\end{equation}
where we have gone to Fourier space using that $\bar{c}^{\mu}(x) = \int_p e^{-i p \cdot x} \bar{c}^{\mu}(p)$ and $c^{\mu}(x)= \int_p e^{i p \cdot x} c^{\mu}(p)$. In general, the induced action of course depends on covariant derivatives with respect to the full and the background metric, but here it suffices to evaluate it in a single-metric approximation $g_{\mu \nu}= \bar{g}_{\mu \nu}$ and on a flat background $\bar{g}_{\mu \nu}= \delta_{\mu \nu}$.  When fixed-point values of couplings are nonvanishing in this approximation, there is clearly no way for them to be zero in the more general case.

The vertex function $V_{\mu \nu}(p_1,p_2,p_3)$ comprises different tensor structures at fixed power of momenta, such as $\bar{c}^{\mu}\partial^2 c_{\mu} \phi \partial^2 \phi$, $\partial_{\nu}\bar{c}^{\mu} \partial_{\nu}c_{\mu} \partial^{\kappa}\phi \partial^{\kappa}\phi$ etc.  For our purpose, it is not important to disentangle the flow of these contributions. To investigate, whether ghost-matter interactions are induced at all, it suffices to study the flow of V in the above approximation. The $\beta$ function of the sum of couplings, that we project on, can only show a Gau\ss{}ian fixed point, if each of the separate couplings has a Gau\ss{}ian fixed point.
Thus we will project onto the simplest nonvanishing component by evaluating the induced flow of
\begin{widetext}
\begin{equation}
\bar{v}(k):=\frac{1}{4 \cdot 48} \left(\left(\frac{\partial^2}{\partial q_{\mu} \partial q^{\mu}}\right)^2 \delta_{\alpha \beta}\left(\int_{q_4}\frac{\delta}{\delta \bar{c}^{\alpha}(q_3)} \frac{\delta^2}{\delta \phi(q_1) \delta\phi(q_2)} \Gamma_k \frac{\overset{\leftarrow}{\delta}}{\delta c^{\beta}(q_4)}\right)\Big|_{q_1=q_2=q_3=q}\right)\Big|_{\phi=0, c=0,q=0}.
\end{equation}
\end{widetext}

Lower orders in the external momentum vanish, as is in accordance with the fact that the ghost-antighost-graviton vertex depends on the momentum of the ghost, and the scalar-squared graviton vertex depends on the momentum of the scalar, see app.~\ref{vertsandprops}.
Thus no ultralocal interaction term is generated, instead the generated operators are momentum-dependent and operators with arbitrarily high powers of momenta will exist. This might be understood as a specific form of (mild) non-locality in the UV. In a standard setting it is known that, starting from a local microscopic action, integrating out quantum fluctuations towards the IR yields nonlocal terms. This differs in the case of asymptotic safety: Diagrams which generate, e.g., matter-ghost couplings with arbitrarily many derivatives, are nonzero as soon as metric fluctuations exist. Thus momentum-dependent ghost-couplings as well as matter couplings \cite{Eichhorn:2012va} will be nonzero at the fixed-point. Accordingly in the case of asymptotic safety the fixed-point action itself seems to be nonlocal in this way. 
Note that this is a mild form of nonlocality, where terms such as $\frac{1}{D^2}$ are not included as separate operators in theory space (note that they could still arise from a resummation of local operators in the IR limit, see, e.g., \cite{Codello:2010mj}). Whether it is actually necessary to extend theory space to include such strongly nonlocal operators is ultimately an experimental question.
%The type of nonlocality appearing here already yields rather complicated effective equations of motion, and could provide for a way to preserve unitarity beyond the perturbative regime, since operators with arbitrarily high powers of derivatives are expected to appear.
On the other hand, it might also be possible that all these operators can actually be resummed to give a very simple expression, for the fixed-point action see, e.g., \cite{Benedetti:2012dx} for evidence of such a scenario.

We now define the dimensionless couplings
\begin{eqnarray}
v(k)&=& \frac{\bar{v}(k) k^4}{Z_c Z_{\phi}},\nonumber\\
g(k)&=& \frac{G_N k^2}{Z_N},\nonumber\\
\lambda(k)&=& \frac{\bar{\lambda}(k)}{k^2},
\end{eqnarray}
and the anomalous dimensions
\begin{eqnarray}
\eta_N&=& - \partial_t \ln Z_N,\nonumber\\
\eta_{\phi}&=& - \partial_t \ln Z_{\phi},\nonumber\\
\eta_c &=& - \partial_t \ln Z_{c}.
\end{eqnarray}
Then the $\beta$ function for $v$ will have the following form
\begin{equation}
\beta_v = 4 v + \eta_c v +\eta_{\phi}v + c_1 g^2 f_1(\lambda) + \mathcal{O}(v).
\end{equation}
Here, $c_1$ is a regularization-scheme dependent constant and $f_1(\lambda)$ is a scheme-dependent function of the cosmological constant.
For $c_1 \neq 0$, $v=0$ is \emph{not} a fixed point, instead the Gau\ss{}ian fixed point is shifted to an interacting one.
To check whether this is the case, we only need to calculate the contribution $\sim c_1 g^2$ to the $\beta$ function.
As a first result, we report this contribution to $\beta_{\bar{v}}$ for a generic regulator:
\begin{eqnarray}
&{}&\partial_t \bar{v}(k)\Big|_{\bar{v}=0}\nonumber\\
&=&\!\! \frac{1}{2 \cdot 4 \cdot 48} \Bigl( \frac{Z_{\phi}Z_c^2}{3}\sqrt{2}(-720) \tilde{\partial}_t \Bigl[\int \frac{d^4p}{(2 \pi)^4} \frac{p^2}{ \mathcal{P}_{k\, TT}^2(p) \mathcal{P}_{k\, \bar{c}c}(p)}\Bigr]\nonumber\\
&-& \frac{1}{4} Z_{\phi}^2 Z_c^2\frac{17}{\sqrt{2}}\tilde{\partial}_t \Bigl[\int \frac{d^4p}{(2 \pi)^4} \frac{p^4}{ \mathcal{P}_{k\, h}^2(p) \mathcal{P}_{k\, \phi}(p)\mathcal{P}_{k\, \bar{c}c}(p)}\Bigr]
\Bigr)\nonumber\\
&{}& = c_1 g^2 f_1(\lambda) \frac{Z_c Z_{\phi}}{k^4}.\label{vfloweq}
\end{eqnarray}
In this expression $\mathcal{P}_{k\, \Phi}$ denotes the regularized inverse propagator for the field $\Phi = h^{TT}, h, \phi, \bar{c}, c$, see app.~\ref{vertsandprops}.
Herein the factor $\frac{1}{4 \cdot 48}$ arises from our definition of the coupling which is motivated by the fact that $\left(\frac{\partial^2}{\partial q_{\mu} \partial q^{\mu}}\right)^2 (q^2)^2 = 4 \cdot 48$. 

This expression shows that the ghost-scalar coupling will be nonzero as soon as metric fluctuations are taken into account: Every metric propagator $\mathcal{P}_{k\, TT}^{-1}$ and $\mathcal{P}_{k\, h}^{-1}$ comes with a factor of $G_N$, and the momentum-integrals over the scale-derivative of the propagators are non-vanishing, thus the right-hand side of \Eqref{vfloweq} is nonzero.
In our case, the factors responsible for this result depend on $G_N$, but in fact in the case of a higher-derivative fixed point action, the corresponding terms would simply be proportional to the higher-derivative couplings. This is evident from \Eqref{vfloweq}, which depends on the regularized graviton propagator, and is therefore nonvanishing for the Einstein-Hilbert action as well as any type of higher-derivative gravitational action.

\begin{figure}[!here]
\includegraphics[width=0.9\linewidth]{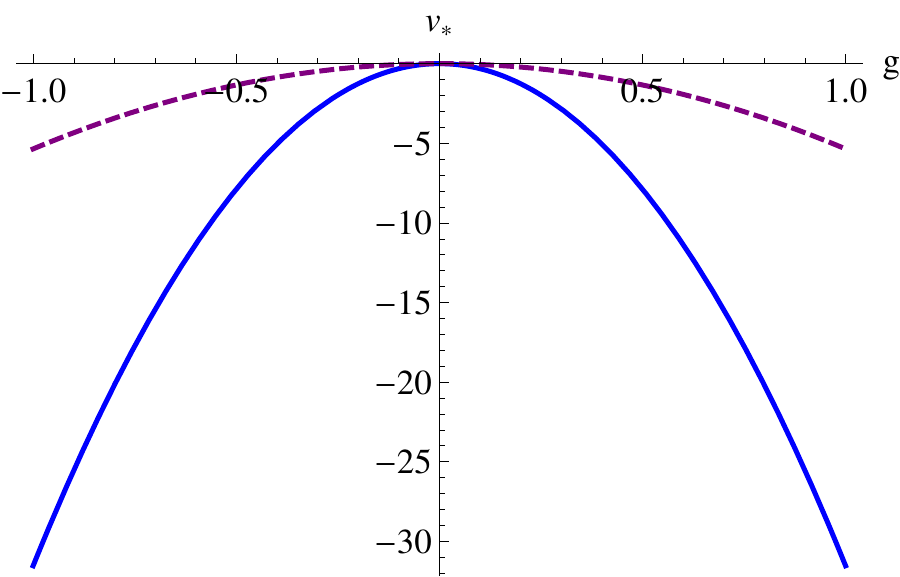}
\caption{\label{vFP} Here we plot the fixed-point value of $v$ as a function of $g$, for $\eta_{N}=-2$, $\eta_{\phi}=-0.78$, cf. \cite{Eichhorn:2010tb}, for a regulator of the form \Eqref{thetacutoff}. The blue thick curve shows the value for $\lambda=0$, whereas the dashed purple curve shows the result for $\lambda=-0.5$.}
\end{figure}

The transition to the $\beta$ function for the dimensionless coupling $v$ then works as follows:
\begin{equation}
\partial_t v(k) = 4 v(k) + \eta_c v(k)+  \eta_{\phi} v(k)+ k^4\frac{\partial_t \bar{v}(k)}{Z_c Z_{\phi}}.
\end{equation}
As pointed out, \Eqref{vfloweq} implies that $k^4\frac{\partial_t \bar{v}(k)}{Z_c Z_{\phi}} \sim g(k)^2$, due to the square of the metric propagator. Thus we observe that $\partial_t v(k)  = 4 v(k) + \eta_c v(k)+  \eta_{\phi} v(k) + c_1\, g(k)^2f_1(\lambda)$, with $f_1(\lambda) \neq 0$ for any finite $\lambda$ and $c_1 \neq 0$. Accordingly, the fixed-point value of $v(k)$ will depend on the value of $g$ quadratically  in this approximation, see fig.~\ref{vFP}.

In the following, we choose a regulator of the form \cite{Litim:2001up}
\begin{equation}
R_k = \left(-\Gamma_k(p^2)+ \Gamma_k(k^2)\right) \Theta(k^2 -p^2)\label{thetacutoff}
\end{equation}
to arrive at the numerical results in fig.~\ref{vFP}. In the pure Einstein-Hilbert truncation with standard Faddeev-Popov ghost term \cite{Eichhorn:2010tb}, as well as in different truncations taking into account the back-coupling of the scalar \cite{Percacci:2002ie, Eichhorn:2012va}, $g_{\ast}>0$ and $\lambda_{\ast}>0$. This implies that $v_{\ast} \neq 0$, and confirms our expectation that metric fluctuations remove the Gau\ss{}ian fixed point in the ghost-matter coupling.\\

Here we point out for the first time that nontrivial ghost-matter interactions exist in asymptotically safe quantum gravity, see fig.~\ref{vFP}.  Note that since vertices coupling two gravitons to two gauge fields and two fermions exist as soon as kinetic terms for those fields are included in the truncation, one should expect that ghost-gauge-field or ghost-fermion interactions will be generated. In fact, the fixed-point action will presumably contain nonzero couplings for operators of the form $\mathcal{O}_g(g_{\mu \nu}) \mathcal{O}_m(m) \mathcal{O}_{c}(\bar{c},c)$, where $\mathcal{O}_{g/m/c}$ denotes operators depending on the metric, matter fields and ghosts, respectively. This effect has already been pointed out for the case of fermions in \cite{Eichhorn:2011pc}.

\subsection{Ghost self-interactions}
\begin{figure}[!here]
\includegraphics[width=0.4\linewidth]{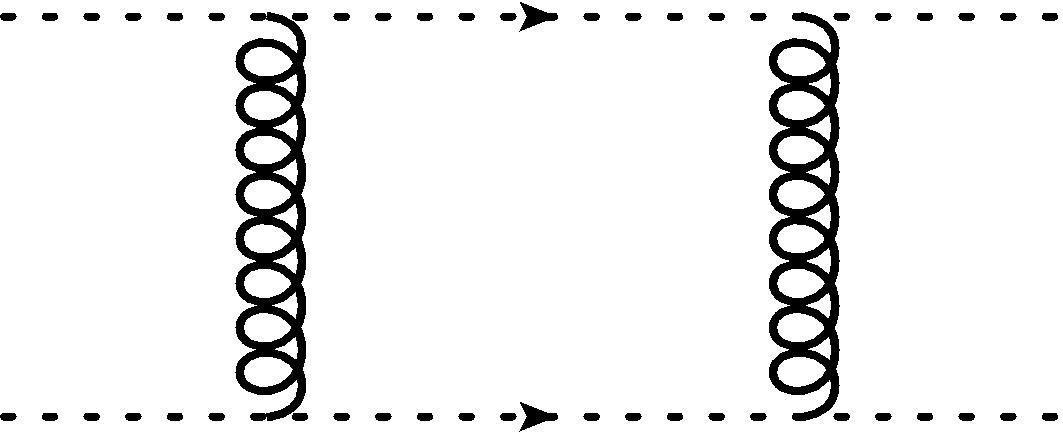} \quad
\includegraphics[width=0.4\linewidth]{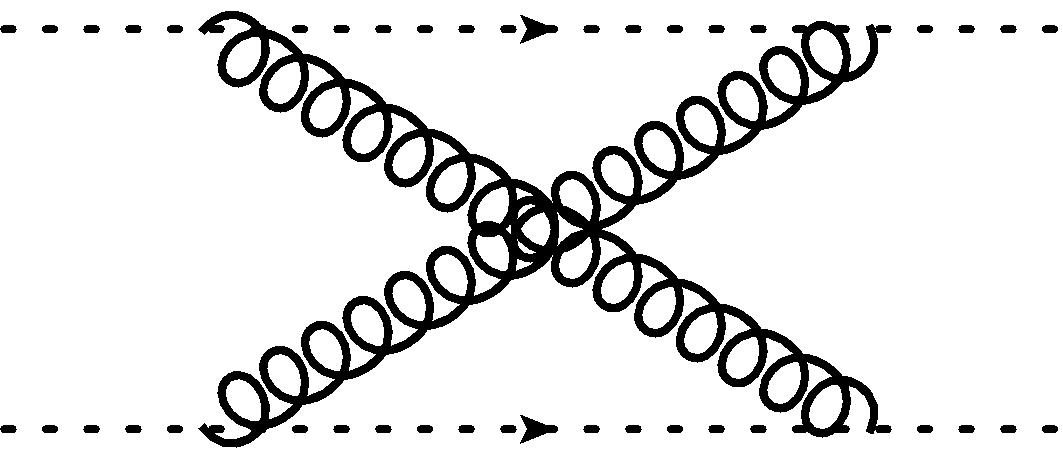}
\caption{\label{ghostdiags} These are the only two diagrams that induce four-ghost couplings, starting with a simple perturbative FP term in the action. Regulator insertions can be found on each of the internal lines.}
\end{figure}

In the following we will consider the truncation \Eqref{truncation} and set the matter action to zero, to study the generation of ghost self-couplings. In the case of the standard Faddeev-Popov ghost term, only a ghost-antighost-graviton vertex exists, and no vertex with coupling to several gravitons. Thus the only diagrams inducing ghost self-interactions are four-vertex diagrams. 
Here, we evaluate the ${\rm Str}\left(\mathcal{P}^{-1}\mathcal{F}\right)^4$ contribution, projected on terms with two external ghosts and two antighosts, see fig.~\ref{ghostdiags}. We first observe, that these diagrams induce a momentum-dependent interaction,  since the ghost-antighost-graviton vertex depends on the momentum of the ghost and vanishes if it is taken to zero. Note that, unlike in \cite{Eichhorn:2011pc}, no cancellation between ladder and crossed-ladder contributions occurs here and in the case of ghost-matter interactions, which would only hold in the case of constant external fields. 

\begin{widetext}
We project the flow equation onto the following ghost self-coupling
\begin{equation}
\bar{\chi}_{\rm gh}= \Bigl(\frac{1}{4\cdot 48}\left(\frac{\partial^2}{\partial q_{\alpha}\partial q^{\alpha}}\right)^2 \left(\int_{q_4} \frac{1}{2}\delta_{\mu \kappa}\delta_{\nu \lambda} \frac{\delta}{\delta \bar{c}^{\mu}(q_4)} \frac{\delta}{\delta \bar{c}^{\nu}(q_2)}\Gamma_k \frac{\overset{\leftarrow}{\delta}}{\delta c^{\kappa}(q_3)}\frac{\overset{\leftarrow}{\delta}}{\delta c^{\lambda}(q_1)}\right)\Big|_{q_1=q_2=q_3=q, \bar{c}=0, c=0} \Bigr)\Big|_{q=0}.
\end{equation}
\end{widetext}

The dimensionless coupling $\chi_{\rm gh}$ is thus given by
\begin{equation}
\chi_{\rm gh} = \frac{\bar{\chi}_{\rm gh} k^4}{Z_{c}^2}.
\end{equation}

Accordingly the $\beta$ function reads
\begin{equation}
\beta_{\chi_{\rm gh}}= 4 \chi_{\rm gh} + 2 \eta_c \chi_{\rm gh} + c_2 g^2 f_2(\lambda)+ \mathcal{O}(\chi_{\rm gh}\cdot g)+ \mathcal{O}(\chi_{\rm gh}^2).
\end{equation}
Herein $c_2$ is a regularization-scheme dependent constant and $f_2(\lambda)$ a regularization-scheme dependent function of the cosmological constant. In the following we will focus on this term in order to point out that for $g\neq 0$, the $\beta$ function cannot have a Gau\ss{}ian fixed point.
 
As in the case of ghost-matter interactions, although our projection does not distinguish different tensor structures, it is fully sufficient to show that the ghost sector has a nontrivial structure beyond a simple perturbative Faddeev-Popov term. Note also that non-unique projections often have to be resorted to in the case of gravity for technical reasons, e.g.,  when employing a spherical background to evaluate the traces on the right-hand side of the flow equation.

We find the following induced $\beta$ function for an unspecified regulator function:
\begin{eqnarray}
\partial_t{\bar{\chi}_{\rm gh}}\Big|_{\bar{\chi}_{\rm gh}=0}\!\!\!\!&=&\!\!\frac{1}{48 \cdot 4} \frac{1}{2} Z_c^4\cdot \nonumber\\
&\cdot&\!\! \Biggl( \frac{-1}{4}\frac{800}{3} \tilde{\partial}_t\int\!\! \frac{d^4p}{\left( 2 \pi\right)^4} \frac{p^4}{\left(\mathcal{P}_{k\, c\bar{c}}(p)\right)^2 \mathcal{P}_{k\, h}(p) \mathcal{P}_{k\, TT}(p)}\nonumber\\
&{}&\,\, -\frac{1}{4}\frac{11840}{9}\, \tilde{\partial}_t\int\!\! \frac{d^4p}{\left( 2 \pi\right)^4} \frac{p^4}{\left(\mathcal{P}_{k\, c\bar{c}}(p)\right)^2 \left( \mathcal{P}_{k\, TT}(p) \right)^2}\nonumber\\
&{}&\,\,-\frac{1}{4}(-35) \tilde{\partial}_t\int \!\!\frac{d^4p}{\left( 2 \pi\right)^4} \frac{p^4}{\left(\mathcal{P}_{k\, c\bar{c}}(p)\right)^2 \left( \mathcal{P}_{k\, h}(p) \right)^2}
\Biggr)\nonumber\\
&=& c_2 g^2 f_2(\lambda) \frac{Z_c^2}{k^4}.
\end{eqnarray}

\begin{figure}[!here]
\includegraphics[width=0.9\linewidth]{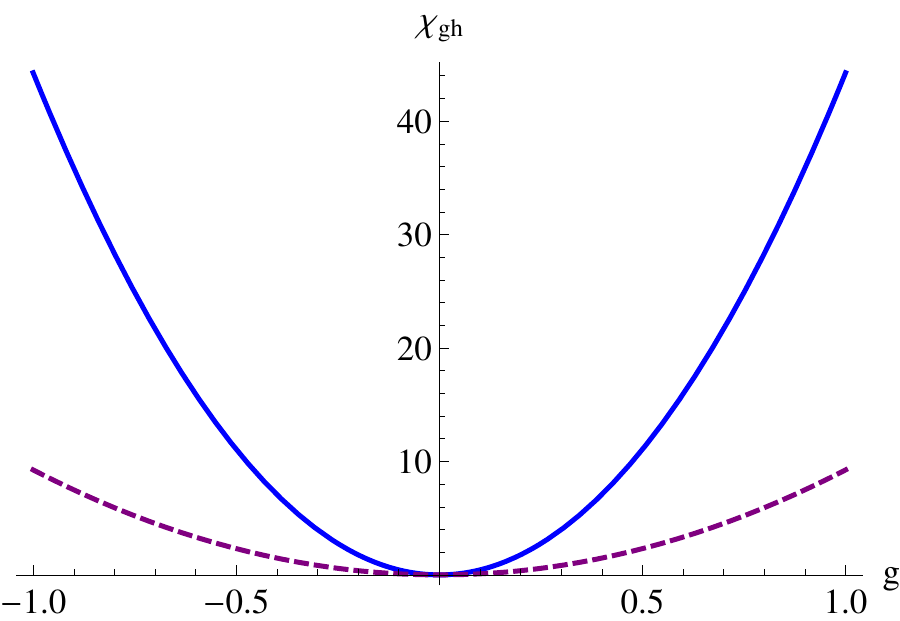}
\caption{\label{four_ghostFP} Here we plot the fixed-point value at the shifted Gau\ss{}ian fixed point for $\eta_c=-0.78$ and $\eta_N=-2$ as a function of $g$ for a regulator of the form \Eqref{thetacutoff}. The blue thick curve shows the result for $\lambda=0$, whereas the purple dashed curve shows the result for $\lambda=-0.5$.
Clearly, the value $\chi_{\rm gh}=0$ can only be reached by setting $g=0$.}
\end{figure}

Herein, the three different terms arise from the York decomposition of the metric field, since the four-vertex diagrams exist with internal transverse traceless or trace modes.
 The four-ghost coupling will be nonzero as soon as metric fluctuations exist, see fig.~\ref{four_ghostFP}. Inserting fixed-point values for $g$ and $\lambda$, which are nonzero in the Einstein-Hilbert and extended truncations, yields $\chi_{\rm gh\, \ast} \neq 0$.
As discussed in the case of ghost-matter interactions, the specific form of the graviton propagator is not important for this effect to exist, and any form of higher-derivative gravitational action will also show the existence of ghost self-interactions.

\section{Discussion and summary: Ghost sector of asymptotically safe quantum gravity}

We have shown that ghost-matter couplings and ghost self-couplings are induced by metric fluctuations. Their $\beta$ functions do not admit a Gau\ss{}ian fixed point if the gravitational couplings are nonvanishing. Thus, the fixed-point action for quantum gravity contains nonvanishing matter-ghost operators and higher-order ghost operators.

The matter-ghost couplings have an interesting implication for the gauge-fixing: Since these are of second order in the ghosts, one can re-express the ghost action in terms of a Faddeev-Popov determinant. Thereby the determinant becomes explicitly dependent on the matter fields, implying a matter-dependent form of gauge fixing. This is reminiscent of the idea to use matter fields, specifically interaction-less 'dust', to introduce a preferred time-slicing and therefore a gauge fixing in Loop Quantum Gravity \cite{Brown:1994py,Rovelli:1993bm}. 

The existence of four-ghost couplings implies that writing the ghost sector as a determinant in the path-integral over metric fluctuations is not possible, although in principle the ghost fields can still be integrated out, even if they occur at higher order. 
Thus, at the interacting fixed point that constitutes the UV completion of the theory, the structure of the theory is very different from the standard setting in gauge theories, where the Faddeev-Popov trick can be reversed and different choices of gauge-fixing and ghost sector are possible.
Here,  the fixed point 'chooses' the structure of the gauge-fixing and ghost sector, and does not seem to be compatible with their perturbative form. The standard way of approaching the quantization of a gauge-theory, where a gauge-invariant action is gauge-fixed, introducing a quadratic ghost term into the path integral seems to break down in the case of asymptotically safe quantum gravity. Instead the fixed point action seems to necessarily make use of a larger number of operators compatible with background-field invariance.

Let us clarify the difference to Yang-Mills theory: There it is known, e.g., from Curci-Ferrari gauges \cite{Baulieu:1981sb} that the most general perturbatively renormalizable BRST invariant action also contains four-ghost operators. Still, there is no need to introduce these terms, as, for instance, Landau gauge without these terms defines a perfectly legitimate choice of gauge. By contrast, asymptotically safe gauge theories, such as gravity, appear to inevitably require higher-order ghost interactions.
Thus a choice of gauge-fixing as part of a truncation of the full effective action that implies the existence of a ghost-antighost-graviton vertex  will show an RG flow that is inconsistent with setting further ghost operators to zero, corresponding to a truncation that is not closed. In other words, the RG flow will generically lead into a region of theory space where higher-order terms in the ghost sector are present and cannot be set to zero consistently in the fixed-point action.
Here, we have shown this with a particular choice of gauge fixing term on the right-hand side of the flow equation, and unspecified regulator shape function. Presumably other choices of gauge-fixing will also exhibit this behavior: One usually constructs a truncation by specifying a gauge-fixing and an accompanying Faddeev-Popov ghost term in addition to the gauge-invariant part of the action.
The gauge-fixing functional $F_{\mu}[\bar{g},h]$ must depend on the background metric and the fluctuations $h_{\mu \nu}$, in order to provide a background-covariant gauge-fixing for the theory. Thus the Faddeev-Popov determinant accompanying this gauge fixing will depend on the fluctuation field. This dependence is enough to ensure that when the determinant is exponentiated with the help of ghost fields, a ghost-antighost-graviton vertex exists. From this vertex, diagrams such as those in fig.~\ref{ghostdiags} can be constructed, and induce nonvanishing higher-ghost operators.
Accordingly the existence of higher ghost operators at the fixed point seems to be a generic feature of asymptotically safe quantum gravity. It seems that the fixed-point action cannot be considered as a diffeomorphism invariant part accompanied by a standard gauge-fixing and ghost term, but exhibits further nonvanishing ghost operators.
%Instead the ghost sector is considerably more complicated and does not allow to simply integrate out the ghost fields to rewrite the ghost action in the form of the Faddeev-Popov determinant, which relies on the quadratic occurrence of the ghosts. 
In this sense, the fixed-point action is very different from a standard classical, i.e., microscopic action, where such a procedure is always possible\footnote{As has been noted in \cite{Manrique:2008zw}, the transition from the fixed-point action $\Gamma_{k \rightarrow \infty}$ to a microscopic action $S$ is nontrivial, and it remains to be investigated what the implication of the nontrivial structure in the ghost sector for this transition is.}. 
Thus at an interacting fixed point, the ghost sector is necessarily more involved than in the setting of an asymptotically free gauge theory, and at the microscopic level the quantum theory only exists in a gauge-fixed version.

Note that RG flows based on the geometric or Vilkovisky DeWitt effective action, such as first studied explicitly in \cite{Donkin:2012ud} clearly are highly interesting in the non-perturbative setting, since a scenario such as the one discussed here could possibly be avoided there.

\subsection{Fixed-point requirement for ghost couplings}

 In the asymptotic-safety scenario observables in an effective theory stay finite, even if the cutoff scale of the effective theory is taken to infinity and the theory becomes fundamental. This holds if all (dimensionless) couplings that enter observables independently approach a finite fixed-point value in the UV (for other possibilities of UV completions see, e.g., \cite{Litim:2012vz,Bonanno:2012dg,Eichhorn:2012uv}). As couplings of ghost operators cannot be measured in experiments, one might conclude that accordingly there need not be a fixed-point requirement for these couplings.
For matter-ghost couplings this is actually different, since matter couplings enter observable quantities. Thus all matter couplings should approach finite fixed-point values. The matter-ghost coupling that we have studied here directly enters matter $\beta$ functions due to the quadratic occurrence of the ghosts. Thus, we face a situation where $\beta_{g_m} \sim v$ and further ghost couplings, for matter couplings $g_m$. Accordingly matter couplings will not stay finite in the UV, if ghost-matter couplings do not approach a fixed point.  A similar consideration actually applies to ghost-curvature couplings which are quadratic in the ghosts. Taken together this implies that all ghost couplings, also higher-order ones, should approach finite fixed-point values for the asymptotic-safety scenario to be viable.

Still, a different scenario is possible:
In principle, taking into account the modified Ward-identities will lead to restrictions on the ghost couplings, relating them to unphysical ("longitudinal") metric couplings. Thus, a divergence of a ghost coupling could be cancelled by the divergence of an unphysical metric coupling, yielding finite predictions for physical observables. This option clearly deserves to be investigated further. Note however that if this option was realized, it would point to a major difference between gravity and Yang-Mills theory: The latter shows an IR-divergent ghost propagator, e.g., in Landau gauge, which is not accompanied by a corresponding behavior of the gluon propagator, \cite{von Smekal:1997is,Fischer:2006vf, Fischer:2009tn}.

\subsection{Relevant couplings in the ghost sector}\label{relevantcoupling}
Let us address the question of relevant couplings in the ghost sector: At the shifted Gau\ss{}ian fixed point in our approximation, the critical exponents are given by
\begin{eqnarray}
\theta_{v}&=& - \frac{\partial \beta_v}{\partial v} = -4- \eta_c- \eta_{\phi},\nonumber\\
\theta_{\chi_{\rm gh}}&=& - \frac{\partial \beta_{\chi_{\rm gh}}}{\partial \chi_{\rm gh}}=- 4 -2 \eta_c.
\end{eqnarray}
Accordingly, for $\eta_c<0$, as observed in the truncation  in \cite{Eichhorn:2010tb} and \cite{Groh:2010ta}, the two couplings are shifted towards relevance, but remain irrelevant for the value of $\eta_c$ in the Einstein-Hilbert truncation. A positive value of $\eta_{\phi}$, as found in \cite{Eichhorn:2012va}, shifts the ghost-matter coupling further into irrelevance. 
Beyond our truncation, also ghost operators of canonical dimensionality 0 and -2 are generated, which will probably be shifted into relevance, see \cite{Eichhorn:2009ah,Eichhorn:2011gc}.

The interpretation of such relevant couplings in the ghost sector is challenging since
 relevant couplings correspond to free parameters, the values of which need to be fixed before the IR value of other couplings is determined. For the coupling of a metric operator, one can hope to find a connection to an observable quantity (at least in principle), such that an appropriate experiment could fix the value of this coupling at some scale. Such a procedure seems impossible for an operator containing a ghost. On the other hand, the RG flow does not 'know' about the distinction between physical and unphysical fields: To uniquely determine a trajectory in theory space predicting the values of all irrelevant couplings and fixing the physics content of the theory in the IR, all relevant couplings need to be assigned a value. Thus, the IR theory remains undetermined as long as relevant couplings in the ghost sector are not fixed. There seem to be two ways to solve this apparent problem:
In the first case, different values of the relevant ghost couplings correspond to distinct physical theories. This case would contradict our understanding of the role of ghosts, and would imply that ghost fields do more than cancelling the effect of unphysical metric components, but can be combined into operators which are accessible to physical measurements\footnote{Yang-Mills theory in the IR seems to provide an example for a theory where quadratic ghost operators contribute to physical observables, such as, for instance, deconfinement order parameters \cite{Braun:2007bx}. The crucial point here is that although ghost operators add important contributions to the calculation of physical observables, their function is the cancellation of unphysical gauge modes. Crucially, no free parameter is associated with any ghost operator in Yang-Mills theory. Thus, if, e.g., higher-order ghost operators carried leading contributions to physical observables, the value of their coupling could be \emph{calculated} from the knowledge of the relevant coupling in Yang-Mills theory, which is directly accessible to measurements. In the case of a relevant ghost coupling, this would be different, since the values of ghost couplings as well as gauge couplings could \emph{not} be calculated from the knowledge of relevant gauge couplings.}.
To understand this scenario it is helpful to first integrate out ghost fields in the path-integral, which is possible  in principle even if these occur at higher order. The resulting path-integral over metric configurations does not take the form of an exponentiated local action any more. Still this could provide a way to identify the free parameters connected to relevant ghost couplings with prefactors of metric operators. Re-exponentiating the new terms would result in a non-local form of the action. Relevant terms in the ghost sector therefore suggest the existence of relevant non-local operators. Alternatively, a free parameter could also exist in a non-trivial measure factor in the path-integral.

In the second case, different values of the relevant ghost couplings correspond to RG trajectories which differ in the values of (some) couplings, but agree in all physical predictions. This is possible if the distinction between RG trajectories due to the relevant ghost couplings is not a physical distinction, but arises from our inability to parameterize the system in terms of physical (and presumably nonlocal) degrees of freedom only.

In this connection it should be mentioned that it could be possible to reformulate the fixed-point action in terms of other fields, in which the distinction between physical and unphysical degrees of freedom is clearer. As an example,  consider QCD, where it is advantageous to introduce auxiliary fields in the IR, using  bosonization techniques \cite{Gies:2001nw}. The RG flow then generates dynamics for these fields and thus turns them into physical fields, which can be identified with mesons, see also \cite{Braun:2011pp}. Similarly, it might be possible to map the fixed-point action in gravity to a different action in terms of other fields, where the distinction of physical and unphysical degrees of freedom ist more transparent, and only physical fields can enter relevant operators.

Clearly, to suggest a solution to the issue of relevant couplings in the ghost sector requires the knowledge of true physical observables in (quantum) gravity.  

A final possibility could be the existence of an IR attractive fixed point, see \cite{Donkin:2012ud,Nagy:2012rn,Christiansen:2012rx}, the domain of attractivity of which comprised all values of the relevant ghost couplings. Since the effective descriptions provided by $\Gamma_k$ must lie on a line of constant physics, the independence of the full effective action $\Gamma_{k\rightarrow 0} = \Gamma_{\ast\, IR}$ from the relevant ghost couplings implies that the distinction of different trajectories $\Gamma_{k>0}$ by different values of these couplings does not have any imprint on observable quantities.

\subsection{Gribov problem and non-perturbative structure of the ghost sector}
Let us discuss the (in)famous Gribov problem: It can arise if the Faddeev-Popov trick, devised to deal with a gauge theory in the perturbative regime, is applied also beyond: As an example, consider  Yang-Mills theory in the Landau gauge, see also \cite{Fischer:2006ub,Maas:2011se}: The Faddeev-Popov operator is given by $- \partial_{\mu}\mathcal{D}_{\mu}^{ab}$, where $\mathcal{D}$ denotes the covariant Yang-Mills derivative and $a,b$ denote indices in the adjoint representation. Whereas this operator remains positive-definite in the perturbative regime, its lowest eigenvalue changes sign at the (first) Gribov-horizon \cite{Gribov:1977wm,Zwanziger:1993dh}, where the value of the gauge field is larger. This happens since the Landau gauge does not uniquely specify a physical field configuration. Accordingly the derivative of the gauge-fixing functional along a gauge orbit, the Faddeev-Popov operator, cannot stay positive definite.
In the non-perturbative regime, the Faddeev-Popov trick for covariant gauges does therefore \emph{not} correspond to inserting a "1" into the functional integral, instead one inserts a "0", making the functional integral ill-defined, \cite{Singer:1978dk}. In gravity, the Gribov problem has been discussed in \cite{Das:1978gk,Esposito:2004zn,Anabalon:2010um}. We observe an interesting alteration to the standard problem in our setting: 
At locations in metric configuration space where the simple FP determinant would be zero, the additional terms in the ghost action, which will in general depend on the background metric and the matter fields, need not be zero.
Thus the location of the Gribov horizon is shifted. 
 Whether it is even possible to completely remove the Gribov horizon(s) remains to be investigated by explicitly studying the lowest-lying eigenvalues of the Faddeev-Popov (FP) determinant in field configuration space. 
 It would indeed be very exciting, if the theory would find a solution to the Gribov problem in the non-perturbative regime "by itself", by requiring the existence of further ghost couplings at the fixed point. 
Studying the ghost action in metric configuration space in our truncation could provide a first indication of whether the Gribov problem is absent in untruncated theory space.

\subsection{Comparison: Ghost sector in asympotically free vs. safe theories}
\begin{figure}[!here]
\includegraphics[width=0.48\linewidth]{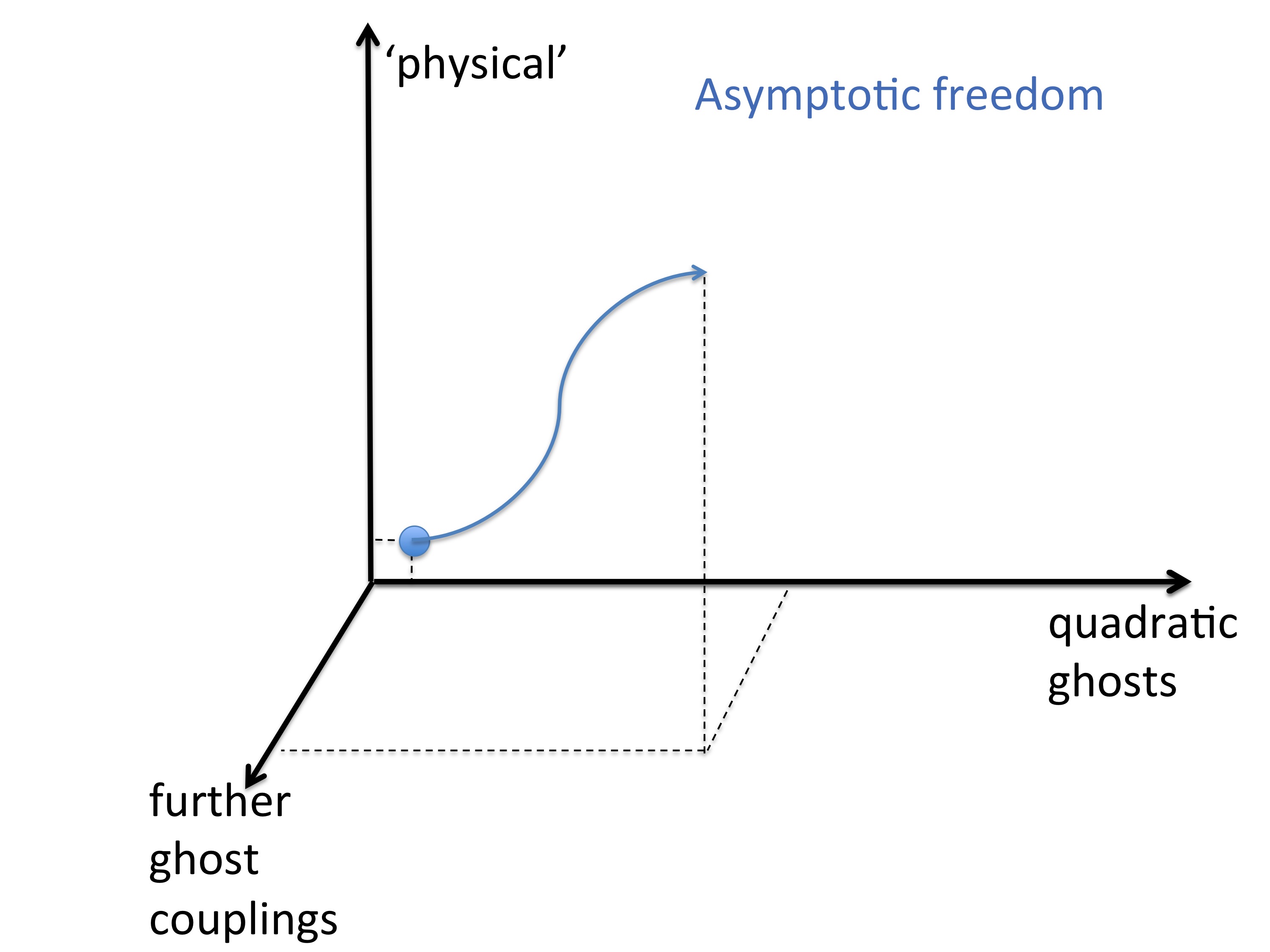}\quad \includegraphics[width=0.48\linewidth]{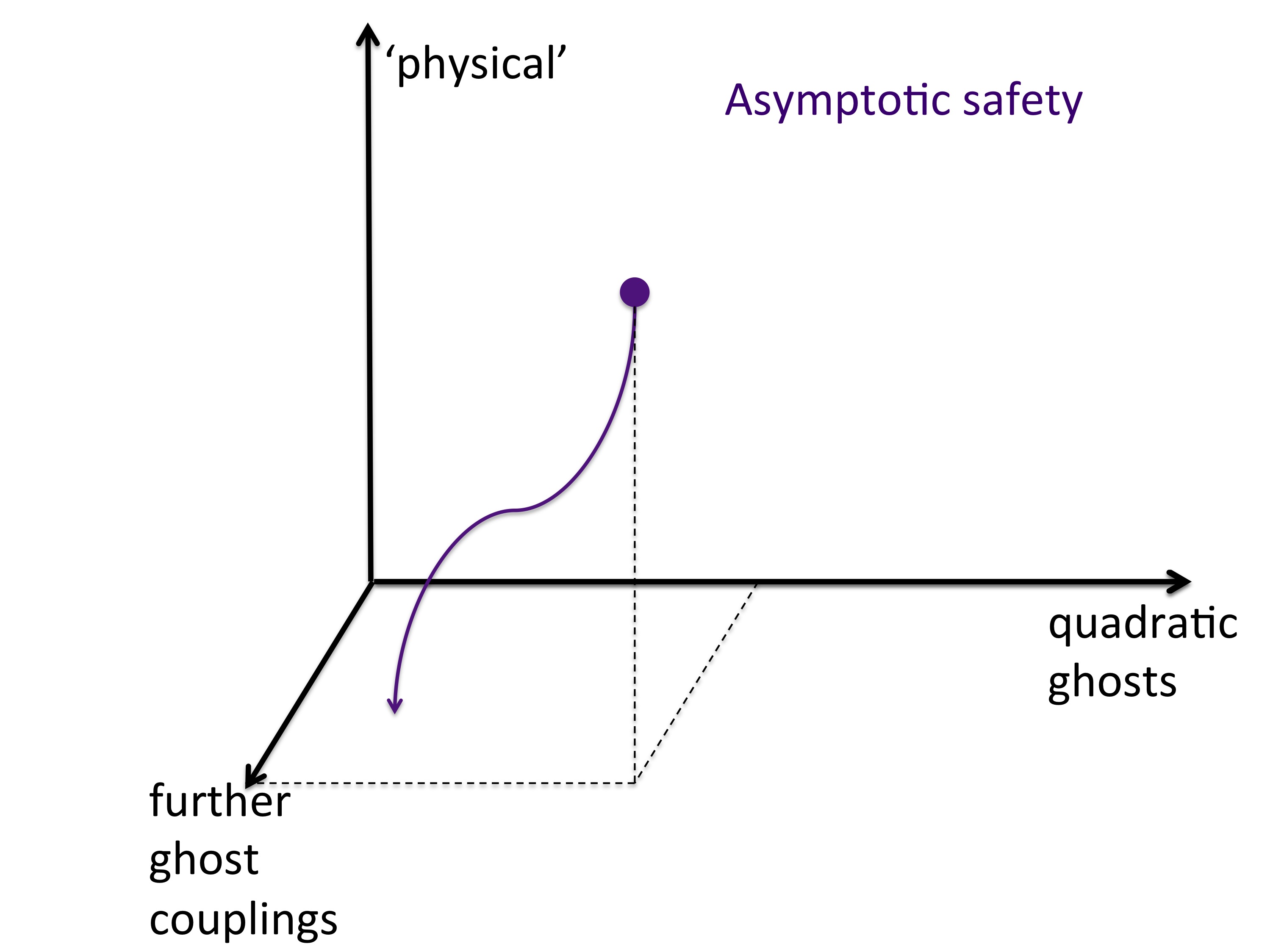}
\caption{\label{AF_AS_comp} We illustrate the RG flow in theory space in the case of asymptotic freedom (left panel) and asymptotic safety (right panel).}
\end{figure}

To clarify the structure of the ghost sector and its implications,
let us point out the difference between asymptotically free gauge theories which become strongly-interacting in the IR, such as Yang-Mills theory, and a non-perturbative UV completion for gravity, see fig.~\ref{AF_AS_comp}.

In both cases, the non-perturbative regime shows a nontrivial ghost sector: In the case of Yang-Mills theories, one can start from a very simple form for the microscopic action with a standard Faddeev-Popov ghost sector in the UV. Gluonic fluctuations will then generate effective ghost-interactions in the IR. In fact, in some gauges the ghosts even become dynamically enhanced \cite{von Smekal:1997is,Fischer:2006vf, Fischer:2009tn} and carry important physical information on, e.g., confinement, see, e.g., \cite{Pawlowski:2010ht}.
Thus at a first glance the situation looks similar to the case of asymptotically safe quantum gravity. The crucial distinction lies in the difference between asymptotic freedom and asymptotic safety: In the former, the UV fixed-point action shows a perturbative ghost sector, whereas further ghost couplings are present in the UV in the second case. 
Besides, only the gauge coupling is marginally relevant in Yang-Mills theory, and none of the ghost couplings is.  Thus ghost-self couplings do not correspond to free parameters in the theory, although they are generated non-perturbatively. The distinction between physical degrees of freedom and ghosts is very clear in this setting: No physical quantity can actually depend on a free parameter in the ghost sector. Furthermore, the choice of gauge is a freedom of the theory, and, e.g., in the case of lattice simulations it can be advantageous to avoid any gauge fixing and simply work with the gauge-invariant microscopic action.
In contrast, the UV fixed-point action in asymptotically safe quantum gravity is nontrivial in the ghost sector, and it is not possible to write the action in terms of a simpler gauge-invariant action by reversing the Faddeev-Popov trick. Besides, relevant couplings in the ghost sector suggest that this sector might even carry free parameters of the physical theory.
In summary, the ghost sector plays a different role in gravity, being crucial for the microscopic definition of the theory.

\subsection{Ghost sector in an effective-field theory setting for gravity}
Our results are not restricted to the case of asymptotically safe quantum gravity. In a more general context, they apply in the effective-field theory framework for quantum gravity, see \cite{Burgess:2003jk,Donoghue:1993eb}. An important difference arises since in that context one could possibly set the microscopic values of the new couplings to zero, since there is no fixed-point requirement at the microscopic scale (which is finite in the effective-field theory setting). Instead the underlying UV completion determines the values of the couplings at this scale, and it is conceivable that the ghost sector could be trivial in such a setting. Since the microscopic theory is defined in a different way, the challenges arising from the fixed-point setting considered here might not carry over to the effective theory.
Then the couplings investigated here would still be generated in the flow towards the IR, similar to the case of asymptotically free gauge theories, see above.
In this case the generated dimensionless coupling would be small, since $g$ is small in that regime (since the effective theory breaks down where $g \sim \mathcal{O}(1)$.)
Since at the shifted GFP the couplings investigated here remain irrelevant, as suggested by their canonical dimensionality, the value of the dimensionfull couplings would run to zero very quickly. Thus ghost self couplings and matter-ghost couplings could possibly be neglected in the effective-field theory framework for all practical purposes, since their effect on any observable must be very small. 
Note however that depending on the choice of UV completion, a similar situation to the case of asymptotic safety could also arise in other settings.

\section{Outlook: Beyond fourth-order truncations}\label{beyondtrunc}
\begin{figure}[!here]
\includegraphics[width=0.5\linewidth]{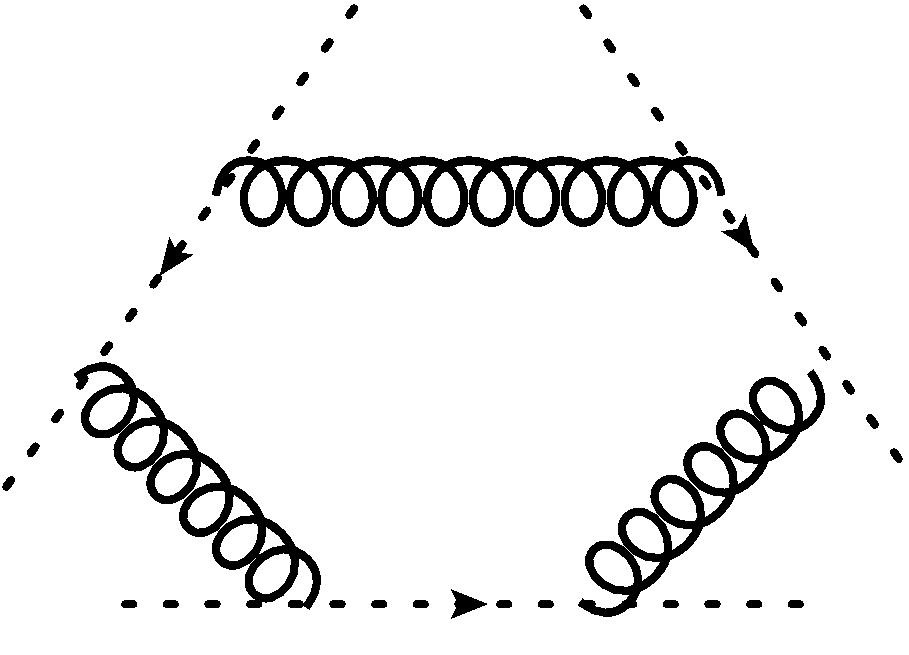}
\caption{\label{higherghost} This six (anti) ghost diagram is constructed from the simple ghost-antighost-graviton vertex, only.}
\end{figure}

So far, we have evaluated the first terms in a presumably infinite number of new ghost couplings. In fact, the ghost-antighost-graviton vertex allows to construct diagrams that induce higher-order ghost couplings (obviously restricted in the maximal number of ghost fields by their Grassmannian nature) of the type 
\begin{equation}
\mathcal{O}_{{\rm gh}\, i}= \int_x\left(\bar{c}^{\kappa} V_{\kappa \lambda}[g_{\mu \nu}, \bar{g}_{\mu \nu}]c^{\lambda}\right)^i,
\end{equation}
see also fig.~\ref{higherghost}.
 Similar diagrams also induce ghost-matter couplings with all matter and gauge fields. Furthermore all these diagrams also induce ghost-curvature-(matter) couplings: Evaluating the flow equation on a curved background, the internal propagators can be derived with respect to the curvature, yielding powers of the curvature in the operator that is induced.
Crucially all these diagrams are generated as soon as a simple Faddeev-Popov term is present in the effective action at some scale. Put differently, the contribution to the $\beta$ function of these new couplings that is generated in this way is independent of the coupling itself. Thereby, setting the coupling to zero does \emph{not} yield a zero in the $\beta$ function. In other words, all these couplings can generically be expected to have a nonzero fixed-point value.
Therefore the structure of the ghost fixed-point action will be very different from a simple Faddeev-Popov ghost sector: Ghosts and matter as well as curvature will be combined into a variety of operators with nonvanishing couplings. 

Many of the new couplings, namely those quadratic in the ghost fields, will directly enter the $\beta$ functions of matter and curvature couplings. Thereby the complicated structure in the ghost sector cannot be ignored, as it enters the flow of couplings that can in principle be linked to observables. Furthermore the question of relevant couplings in the ghost sector becomes more pressing and requires further investigation.
In this context the study of the ghost sector with methods as in \cite{von Smekal:1997is,Fischer:2006vf, Fischer:2009tn} might allow to gain insight into the behavior of the infinite tower of vertex functions involving ghost couplings. 

Judging from the present investigation and that of \cite{Eichhorn:2012va} the following picture seems to emerge: Constructing a fundamental theory of quantum gravity, i.e., a quantum field theory that can exist in the infinite-cutoff limit, with the help of an interacting fixed point, implies that the interactions cannot be contained within a finite number of operators.  Accordingly, the far UV is described by a theory with rather complicated (derivative) interaction terms between all fields in the theory, and the spectrum of quantum fluctuations becomes very involved. Unlike in the case of an asymptotically free gauge theory, where one relevant coupling drives the RG flow, a larger but presumably finite number of such couplings exist, and could also be found in the ghost sector.

In such a setting, numerical simulations of the path-integral for gravity in terms of gauge-invariant degrees of freedom seem to become more challenging. Clearly a direct translation of the fixed-point action to a microscopic action in a discretized setting is not possible. %Nevertheless the existence of the nontrivial structure in the ghost sector and the fact that it is not possible to rewrite the fixed-point action in terms of a gauge-invariant action by simply reversing the Faddeev-Popov trick, seems to pose a challenge for simulations which propose to evaluate the path-integral for gravity in terms of gauge-invariant degrees of freedom only: 
Our investigation seems to imply that at a possible UV fixed point, it is considerably more complicated than in a perturbative setting to write the generating functional without the occurrence of ghost fields by integrating these out. Finally if ghost couplings become relevant, they correspond to parameters that need to be tuned in a discretized setting in order to reach the continuum limit there. As discussed in sect.~\ref{relevantcoupling}, these might actually correspond to parameters in a non-trivial measure factor in the gravitational path integral, possibly related to that in \cite{Laiho:2011ya}, or to a non-local curvature operator.
To summarize, this suggests that the transition from the fixed-point action to a microscopic classical action needs to be investigated further \cite{Manrique:2008zw}, to understand the structure of the ghost sector in this transition. This will help to elucidate the connection between simulations such as those in \cite{Ambjorn:2012jv} and the present setting of the FRG.  It might actually be possible that a theory formulated in terms of physical degrees of freedom only, in fact lies in a different universality class than one which employs ghost fields and contains relevant couplings in the ghost sector. 

In the future, it is mandatory to investigate infinite-dimensional truncations, e.g., functions of the operators considered here. One might hope that the asymptotic form of these functions in the far UV becomes simple, as advocated in \cite{Benedetti:2012dx} for the case of curvature operators. Otherwise the structure of the theory as implied by the present investigation and that in \cite{Eichhorn:2012va} seems to suggest that tools complementing the FRG approach to gravity should be developed in order to get a handle on the complicated structure of the theory.

%
%To summarize, the structure of the ghost sector seems to be the following one: Close to the Gau\ss{}ian fixed point, where we know gravity best, as it corresponds to observable scales, we see a diffeomorphism invariant theory with the Einstein-Hilbert term in the action. To quantize this theory, we then use the path-integral framework which implies the necessity to gauge-fix and introduce a Faddeev-Popov ghost sector. We then observe, that the theory 'makes use' of this sector in a rather nontrivial way: Towards higher scales, the flow does not stay in the part of theory space where the Faddeev-Popov ghost sector is trivial, instead it generates a highly nontrivial ghost sector, with higher-order ghost couplings, ghost-matter couplings, and presumably also ghost-curvature couplings. All these new couplings have $\beta$ functions which do not admit a Gau\ss{}ian fixed point as soon as metric fluctuations are present. Accordingly
%all these dimensionless couplings are nonzero in the UV, and the ghost sector does not resemble a perturbative Faddeev-Popov ghost sector at all.\newline\\

Let us add that the results presented here do not exclude the possibility of the following scenario: Although ghost couplings do not admit a Gau\ss{}ian fixed point, their back-coupling into the flow of operators connected to physical observables could be small. A similar effect has been observed in \cite{Eichhorn:2012va} for a class of matter couplings and their back-coupling into the flow of the Einstein-Hilbert sector, which is in fact subleading. As a point in favor of this scenario, note that a larger number of fermionic matter fields -- as in the standard model -- shifts the fixed-point value of the cosmological constant toward larger negative values \cite{Percacci:2002ie}, which implies that the contribution of metric fluctuations to, e.g., ghost $\beta$ functions is reduced, as discussed in \cite{Eichhorn:2011pc}. Thereby, the fixed-point value at the shifted Gau\ss{}ian fixed point becomes smaller, cf. fig.~\ref{vFP} and fig.~\ref{four_ghostFP}. A smaller fixed-point value in turn implies a smaller back-coupling into the flow of metric operators. 
Furthermore, in the untruncated theory space, all relevant couplings could be connected to physical observables and no ghost coupling would be relevant. In such a setting, the rather complicated structure of the ghost sector would play a subleading role in the calculation of physical predictions from this theory. Whether this scenario, or one with a large back-coupling of ghost operators into the flow of metric couplings, and a finite number of relevant ghost couplings, is actually realized, necessitates more detailed investigations of the ghost sector.\newline\\

{\emph {Acknowledgements}} I would like to express my gratitude for many helpful discussions with Holger Gies and for his interesting comments on this manuscript. I also thank Martin Reuter for useful observations.
I would also like to express my thanks to the numerical relativity group at the University of Illinois at Urbana-Champaign for the hospitality during a part of this work.
Research at Perimeter Institute is supported by the Government of Canada through Industry Canada and by the Province of Ontario through the Ministry of Research and Innovation.\newline\\

\begin{appendix}
\section{Vertices and propagators for the $\mathcal{P}^{-1}\mathcal{F}$ expansion}\label{vertsandprops}
In the following we use a notation where the subscripts $TT$, $h$, $\bar{c}$, $c$ and $\phi$ denote the transverse traceless graviton mode, the trace mode, the antighost, ghost and scalar.

The projection operators for the transverse traceless graviton and the ghost propagator read as follows:
\begin{eqnarray}
P_{TT\, \mu \nu \kappa \lambda}(p)&=& \frac{1}{2}\left(P_{T\, \mu \kappa} (p)P_{T\, \nu \lambda}(p) + P_{T\, \mu \lambda}(p)P_{T\, \nu \kappa}(p)\right) \nonumber\\
&{}&- \frac{1}{3}P_{T\, \mu \nu} (p)P_{T\, \kappa \lambda}(p),
\end{eqnarray}
where $P_{T\, \mu \nu}(p) = \delta_{\mu \nu} - \frac{p_{\mu} p_{\nu}}{p^2}$ denotes the standard transversal projector.
\begin{equation}
P_{\bar{c}c\, \mu \nu} (p)= \frac{1}{\sqrt{2}} \left(\delta_{\mu \nu} - \frac{1}{3} \frac{p_{\mu} p_{\nu}}{p^2}\right).
\end{equation}
Next we define the following vertex functions to facilitate the definition of the elements of the fluctuation matrix involving ghosts and antighosts:
\begin{eqnarray}
V_{T\, \mu \kappa \rho \sigma}(p,q)\!\!&=&\!\!\! \frac{Z_c(k)}{\sqrt{2}} \Bigl( \left(p\cdot q + q^2 \right)\left(\delta_{\mu \rho}\delta_{\kappa \sigma}+ \delta_{\mu \sigma} \delta_{\kappa \rho} \right) \nonumber\\
&{}&+\frac{1}{2} q_{\rho}q_{\mu}\delta_{\kappa \sigma} + \frac{1}{2} q_{\sigma}q_{\mu}\delta_{\nu \kappa} - \frac{1}{2}p_{\mu}q_{\rho}\delta_{\kappa \sigma}\nonumber\\
&{}&- \frac{1}{2}p_{\mu}q_{\sigma}\delta_{\kappa \rho} + p_{\kappa}q_{\rho}\delta_{\mu \sigma} + p_{\kappa}q_{\sigma}\delta_{\mu \rho}\Bigr)\nonumber\\
&{}&\\
V_{\mu \kappa}(p,q)&=&- \sqrt{2}Z_c(k)\Bigl(\frac{1}{4} \left( - p\cdot q - q^2\right) \delta_{\mu \kappa} - \frac{1}{4} p_{\kappa}q_{\mu} \nonumber\\
&{}&- \frac{1}{8} q_{\kappa}q_{\mu}+ \frac{1}{8}p_{\mu}q_{\kappa}
\Bigr).
\end{eqnarray}
Next we define
\begin{eqnarray}
V_{c\,h\, \mu \alpha}(p,q)&=& \bar{c}^{\alpha}(q-p)V_{\mu \alpha}(p,-q)\nonumber\\
V_{\bar{c}\,h\, \mu \alpha}(p,q)&=&-c^{\mu}(p-q)V_{\mu \alpha}(p,q-p)\nonumber\\
V_{h\, c\, \mu \alpha}(p,q)&=&-\bar{c}^{\alpha}(q-p)V_{\mu \alpha}(-q,p)\nonumber\\
V_{h\,\bar{c}\, \mu \alpha}(p,q)&=& c^{\mu}(p-q)V_{\mu \alpha}(-q,q-p)\nonumber\\
V_{c \,TT \, \mu \alpha \gamma \beta}(p,q)&=&- \bar{c}^{\alpha}(q-p)V_{T\, \mu \alpha \gamma \beta}(q,-p)\nonumber\\
V_{\bar{c}\, TT\, \mu \alpha \gamma \beta}(p,q)&=& c^{\mu}(p-q)V_{T\, \mu \alpha \gamma \beta }(q,p-q)\nonumber\\
V_{TT\, \bar{c}\, \mu \alpha \gamma \beta}(p,q)&=&- c^{\mu}(p-q)V_{T\, \mu \alpha \gamma \beta}(-p,p-q)\nonumber\\
V_{TT\, c\, \mu \alpha \gamma \beta}(p,q)&=& \bar{c}^{\alpha}(q-p)V_{T\, \mu \alpha \gamma \beta}(-p,q).
\end{eqnarray}
The full matrix entry of the fluctuation matrix involves an external ghost or antighost field, respectively.
As shown in \cite{Eichhorn:2010tb}, there is no 2-graviton vertex with external ghost and antighost.\newline\\

Finally we have the vertices connecting gravitons and the scalar:
\begin{eqnarray}
V_{\phi\, h}(p,q)&=&- \frac{Z_{\phi}(k)}{4} \phi(p-q)\left( p \cdot q - q^2\right)\nonumber\\
V_{h\, \phi}(p,q)&=& \frac{Z_{\phi}(k)}{4} \phi(p-q)\left(p^2 - p \cdot q \right)\nonumber\\
V_{TT\, \phi\, \mu \nu}(p,q)&=& \frac{Z_{\phi}(k)}{2} \phi(p-q)\left(p_{\mu}q_{\nu}+ p_{\nu}q_{\mu}- 2 q_{\mu}q_{\nu} \right)\nonumber\\
V_{\phi\, TT\, \mu \nu}(p,q)&=& \frac{Z_{\phi}(k)}{2} \phi(p-q)\left( p_{\mu}q_{\nu} + p_{\nu}q_{\mu}- 2 p_{\mu}p_{\nu}\right).\nonumber\\
&{}&
\end{eqnarray}
\begin{eqnarray}
V_{TT\, \mu \nu \kappa \lambda} &=& \frac{Z_{\phi}(k)}{8} \int_{l_1}\phi(l_1)\,\phi(q-p-l_1)\cdot \nonumber\\
&{}& \cdot\Bigl(l_1^{\gamma}(-p_{\gamma}+q_{\gamma}-l_{1\, \gamma})\left(\delta_{\mu \kappa}\delta_{\nu \lambda}+\delta_{\mu \lambda}\delta_{\nu \kappa} \right)  \nonumber\\
&{}&+ \Bigl( \left(l_{1\, \mu}\delta_{\nu \lambda}+l_{1\, \nu}\delta_{\mu \lambda} \right)(p_{\kappa}+l_{1\, \kappa})\nonumber\\
&{}&
+\left(l_{1\, \mu}\delta_{\nu \kappa}+l_{1\, \nu}\delta_{\mu \kappa} \right)(p_{\lambda}+l_{1\, \lambda})\nonumber\\
&{}&
+\left(l_{1\, \kappa}\delta_{\lambda \mu}+ l_{1\, \lambda}\delta_{\kappa \mu} \right)\left (-q_{\nu}+l_{1\,\nu} \right)\nonumber\\
&{}&
+\left(l_{1\, \kappa}\delta_{\lambda \nu}+ l_{1\, \lambda}\delta_{\kappa \nu} \right)\left(-q_{\mu}+l_{1\,\mu} \right)
\Bigr)
\Bigr).
\end{eqnarray}
\end{appendix}

\end{document}